\documentclass[letterpaper,12pt]{article}
\usepackage{amsmath,amsfonts,latexsym}

\newcommand*{\sgn}{\mathop{\rm sgn}\nolimits}
\title{Some solutions of linearized 5-d gravity with brane}
\author{O.~I.~Vasilenko\thanks{E-mail address:
\texttt{oiv@srdlan.npi.msu.su}}
\\[6pt]
\emph{\small Department of Physics, Moscow State University,}\\
\emph{\small  119899 Vorob'evy gory, Moscow, Russia}}
\date{}

\begin{document}
\maketitle \abstract{We consider linearized 5-d gravity in the
Randall-Sundrum brane world. The  class of static solutions for
linearized Einstein equations is found. Also we obtaine wave
solutions  describing radiation from an imaginary point source
located at the Planck distance from the  brane. We analyze the
fields asymptotic behavior and peculiarities of matter sources.}

\section{Introduction and Summary}\label{intro}
For a long time the idea is investigated that the number of
spacial dimensions exceeds three.  This supposition was used in
efforts to unify electromagnetic and gravitational interactions.
String theory predicts an existence of more than three space
dimensions. One interesting attempt \cite{Randall99a,Randall99b}
based on more old ideas \cite{Rubakov83, Akama83, Visser85} is the
proposition to use the higher-dimensional mechanism for solving
the hierarchy problem. It is assumed that the visible flat
$(1+3)$-d spacetime (brane) is embedded within a 5-d spacetime.
Apart from other approaches an additional dimension is considered
to be noncompact. Gravity on the brane resembles the usual
4-dimensional Einstein gravity for long distances due to the
special choice of metric, cosmological constant and brane tension.
The investigation of this scenario shows that  some problems exist
\cite{Mueck00}.

It is important to verify a physical noncontradiction of the
RS-metric: $ds^2=(k|z|+1)^{-2}\eta_{MK}dX^MdX^K$. In this work we
consider small gravitational fields and in particular the
question: is metric perturbations remain small everywhere in the
5-dimensional spacetime? We assume that the brane is spherically
symmetric. By $r/k$ and $z$ denote coordinates on the brane and
normal to the brane respectively. Using variables
$R=\sqrt{r^2+(kz+1)^2}$ and $p=r/R$, we separate linearized
Einstein equations and get an extensive class of static and wave
solutions. Nonstationary solutions depend on time $t$ as
$\exp(i(t-R))$ and describe waves radiating outward from an
imaginary source located at the Planck distance $1/k$ from the
brane. Obtained solutions may be of any degree of decreasing at
infinity. A matter on the brane and stringlike formations lined
orthogonally to the brane represent field sources.

\section{Linearized equations}\label{equations}
The action describing 5-dimensional void spacetime including the
4-dimensional  brane with matter is
\begin{equation}
\label{action}
 S = \int d^5 X\left[\sqrt{|g_{(5)}|}
\left(\frac{R}{2}-\Lambda\right)
 + \sigma\delta(z)\sqrt{|g_{(4)}|}
  (1 +\mathcal{L}_{(4)}^{\text{matter}})\right],
\end{equation}
where $ds^2=g_{MK}dX^MdX^K$ is a 5-dimensional interval of
spacetime with a metric $g_{MK} (M,K=0,1,2,3,5)$ and coordinates
$X^M=(x^\mu,x^5=z), (\mu,\kappa=0,1,2,3)$; $g_{(4)\mu\kappa}$ and
$x^\mu$ are a metric and coordinates on the brane located at
$z=0$.

 5-dimensional Einstein equations are
\begin{equation}
\label{Einstein}
 R_{MK}-\frac{1}{2}g_{MK}R= \Lambda g_{MK}+\sigma \delta(z)
 \left( g_{(4)}^{\mu\kappa}+
T_{(4)}^{\text{matter}\;\mu\kappa}\right)
\sqrt{\left|\frac{g_{(4)}}{g_{(5)}}\right|}g_{\mu M}g_{\kappa K}.
\end{equation}
In the absence of gravitational waves and a matter the solution of
\eqref{Einstein} is given by the background RS-metric
\begin{equation}
\label{RSmetric}ds^2=H^{-2}(z)\eta_{MK}dX^MdX^K,
\end{equation}
where $\eta_{MK}$ is a  Minkowski metric with the signature
$(+,-,-,-,-)$ and $H(z)=k|z|+1$. The cosmological constant
$\Lambda$ and the brane tension $\sigma$ relate to the constant
$k$ as $\Lambda=-6k^2,\sigma=6k$ \cite{Randall99a}.

Consider  background metric perturbations parameterized by a
tensor $h_{MK}$
\begin {equation}
\label{PrtbMetric}g_{MK}=H^{-2}(z)(\eta_{MK}+h_{MK}).
\end{equation}
In the axial gauge
\cite{Randall99a,Randall99b,Garriga99,Giddings00}$ (
h^\kappa_\kappa=0,h_{55}=h_{5\mu}=0)$ linearized Einstein
equations take the form
\begin{equation}
\label{LinEq} \left(-{\Box}_{(4)} + \partial_z^2 - \frac{3\sgn(z)
}{k|z|+1}\partial_z\right)h_{\mu\kappa} =
12k\delta(z)T_{(4)\;\mu\kappa}^{\text{matter}\,(1)},
\end{equation}
where $T_{(4)\;\mu\kappa}^{\text{matter}\,(1)}$ is the 1-st order
contribution of matter on the brane to an energy-momentum tensor
calculated using unperturbed metric. We suppose that there is no
contribution  from a matter on the brane  in zero-order. This
equation comes also from the other gauge choices \cite{Aref'eva}.
Since the operator acting on $h_{\mu\kappa}$ does not depend on
indices $\mu,\kappa$, we shall drop them for simplicity and use
further the definitions $h$ and $T$ for $h_{\mu\kappa}$ and
$T_{(4)\;\mu\kappa}^{\text{matter}\,(1)}$ respectively.

\section{Solution of equations}\label{solution}
Consider first the equation \eqref{LinEq} in the area outside the
brane.  For $z>0$, we have
 \begin{equation}
\label{void}
  \left(-\Box_{(4)} + \partial_z^2 -
 \frac{3k}{kz+1}\partial_z\right)h = 0.
\end{equation}

Suppose the brane is spherically symmetric; then the dependence of
$h$ on $x_1,x_2,x_3$ reduces to the dependence on
$r=k\sqrt{x_1^2+x_2^2+x_3^2}$ only. Let  $t=kx_0,
R=\sqrt{r^2+(kz+1)^2}, p=r/R$ be new variables; then equation
\eqref{void} written for the function $f=rh$ takes the form
\begin {equation}
\label{tRp-eq} -\frac{\partial^2 f}{{\partial t}^2}+
\frac{\partial^2 f}{{\partial R}^2}- \frac{2}{R}\frac{\partial
f}{\partial R}+\frac{1}{R^2}\left((1-p^2)\frac{\partial^2
f}{{\partial p}^2} +2p\frac{\partial f}{\partial p}\right)=0.
\end{equation}
To find a solution we use the method of dividing variables. Let
$f$ takes the form
\begin {equation}
\label{ansatz}
 f=e^{i\epsilon t}R^{3/2}(1-p^2)H(R)P(p),
\end{equation}
where $\epsilon=0,\pm1$. For the function $H$  we obtain the
equation containing an arbitrary constant  $\nu$
\begin{equation}
\label{H-eq}
 R^2H^{\prime\prime}+RH^{\prime}+\left(R^2\epsilon^2
 -\left(\nu+\frac{1}{2}\right)^2\right)H=0.
\end{equation}
If $\epsilon=\pm1$, then \eqref{H-eq} is a Bessel equation. The
function $P$ satisfy the  Legendre equation
\begin {equation}
\label{P-eq} (1-p^2)P^{\prime\prime}-2pP^{\prime}+
\left(\nu(\nu+1)-\frac{4}{1-p^2}\right)P=0.
\end{equation}
A general solution $P(p,\nu)$ of \eqref{P-eq} is a linear
combination of Legendre functions $P^2_\nu(p)$ and $Q^2_\nu(p)$ of
the first and second kinds respectively.

The explicit form of $P(p,\nu)$ for  $\nu=0, 1, 2$ is
\begin{align}
P(p,0)&=c_{01}\frac{p}{1-p^2}+c_{02}\frac{1+p^2}{1-p^2},\notag\\
\label{P(p,nu=0,1,2)}
P(p,1)&=c_{11}\frac{1}{1-p^2}+c_{12}\frac{p-\frac{p^3}{3}}{1-p^2},\\
P(p,2)&=c_{21}(1-p^2) + c_{22}\left(\frac{4p}{1-p^2}+6p+3(1-p^2)
\ln\left(\frac{1+p}{1-p}\right)\right)\notag.
\end{align}
Here and further,  $a_i,c_{ij},$  are arbitrary constants.

 \section{Static solutions}\label{static}
If  we assume $\epsilon=0$ in \eqref{ansatz}, then $h$ does not
depend on time and takes the form
\begin {equation}
\label{Stat}
 h=\left(a_1R^{\nu+1}+a_2R^{-\nu}\right)\frac{1-p^2}{p}P(p,\nu).
\end{equation}
A solution have physical significance if $h$ decreases as $R$
increases, so  $a_1=0$ in \eqref{Stat} for $\nu\geq 0$.

First consider solutions corresponding to the case $\nu=~1$,
\begin{align}
\label{h1(nu=1)}
 h_1&=a_1\frac{1}{r},\\
 \label{h2(nu=1)}
 h_2&=a_2\frac{1}{R}\left(1-\frac{1}{3}\left(\frac{r}{R}\right)^2\right).
\end{align}

In order to find the energy-momentum tensor $T$ corresponding to
$h$, we replace  $z$ by $|z|$ in  $h$ and  act on  $h$ by the
operator from the left part of the equation \eqref{LinEq}.A
nonzero result appears in two cases.

In the fist case solution has a singularity of $1/r$ type as
$r\rightarrow 0$. This is the case of the solution $h_1$. The
result of action on $1/r$ by operator $\frac{1}{r^2}\frac{\partial
}{\partial r}\left(r^2\frac{\partial }{\partial r}\right)$ from
$\Box_{(4)}$ is $\delta(r)$. It signifies that the source of the
field is a string located at $r=0$ along  $z$  direction normally
to brane hyperplane.  The action \eqref{action} does not
presuppose an existence of similar sources however corresponding
local matter term can be easily included. Since initially it was
assumed that perturbations are small, such solutions are correct
in the $r>r_0$ area only. The condition that $h$ is small in
comparison with $\eta$ determines the value of $r_0$. The area
$r<r_0$ is occupied by a source. Notice that similar sources are
typical for the considered geometry, they arises for example  in
nonperturbative solutions of the "black cigar" type
\cite{Chamblin99}.

The second case corresponds to the action of $\partial^2/\partial
z^2$ operator on $h$. Taking into account that $h$ depends on
$|z|$, we have
\begin {equation}
\label{DiffMod} \frac{\partial^2 h}{\partial z^2}=\frac{\partial^2
h}{\partial |z|^2}+\frac{\partial h}{\partial |z|}2\delta(z).
\end{equation}
From \eqref{DiffMod} and \eqref{LinEq} we find $T$
\begin {equation}
\label{TFromZ} T=\frac{1}{6k}\frac{\partial h}{\partial
|z|}\;(z=0).
\end{equation}
Since  $h$ depends on $z$ by means of dependence on $R$, we obtain
\begin {equation}
\label{TFromR} T=\frac{1}{6R}\frac{\partial h}{\partial
R}\;(R=\sqrt{r^2+1}).
\end{equation}
In the case of solution $h_2$ determined by \eqref{h2(nu=1)} we
get
\begin {equation}
\label{T_2} T_2=-\frac{a_2}{6(r^2+1)^{\frac{5}{2}}}.
\end{equation}

As a second example consider  the case $\nu=2$. Solutions take the
forms
\begin{align}
\label{h3(nu=2)}
 h_3&=a_3\frac{1}{rR}\left(1-\frac{r^2}{R^2}\right)^2,\\
 \label{h4(nu=2)}
 h_4&=a_4\left(\frac{10}{R^2}-\frac{6r^2}{R^4}+
 \frac{3}{rR}\left(1-\frac{r^2}{R^2}\right)^2\ln\left(\frac{R+r}{R-r}\right)\right).
\end{align}
Substituting   $h_3$ in \eqref{LinEq}, we receive following
expression for the right part
\begin {equation}
\label{T_3} 2a_3\left(\frac{\delta(r)}{k|z|+1}+
\delta(z)\frac{4r^2-1}{r(r^2+1)^\frac{7}{2}}\right).
\end{equation}
This source consists from  a matter distributed on the brane and a
string. The energy-momentum tensor $T_4$ corresponding to the
solution $h_4$ describes a matter located on the brane only and
looks like
\begin {equation}
\label{T_4} T_4=a_4\left(\frac{4(r^2-5)}{(r^2+1)^3}+
\frac{3(4r^2-1)}{r(r^2+1)^\frac{7}{2}}
\ln\left(\frac{\sqrt{r^2+1}+r}{\sqrt{r^2+1}-r}\right)
-\frac{6}{(r^2+1)^\frac{5}{2}}\right).
\end{equation}

\section{Wave solutions}\label{wave}
Now we assume $\epsilon=1$ in  \eqref{H-eq}. Then \eqref{ansatz}
depends on time and  \eqref{H-eq} has a form of the Bessel
equation. We choose the Hankel function
$H(R)=H_{\nu+\frac{1}{2}}^{(2)}(R)$ as a solution of \eqref{H-eq}.
Then a general solution for $h$ describes outgoing wave. This is
clearly seen in the case  of integer $\nu$ when a simple explicit
expression for $H_{\nu+\frac{1}{2}}^{(2)}(R)$ exists
\begin {equation}
\label{Hankel}
 H_{\nu+\frac{1}{2}}^{(2)}(R) = const\;
 R^{\nu+\frac{1}{2}}\left(\frac{d}{RdR}\right)^{\nu}\frac{e^{-iR}}{R}.
\end{equation}
We present solutions and their asymptotics for some values of
$\nu$.
\paragraph{Case $\nu=0$}
\begin {equation}
\label{h(t,0)}
 h=e^{i(t-R)}\left(c_{01}+c_{02}\left(\frac{R}{r}+\frac{r}{R}\right)\right).
\end{equation}
As $r\to \infty$
\begin {equation}
\label{}
 h\to e^{i(t-r)}\left(c_{01}+
 2c_{02}+\frac{c_{02}}{4}\left(\frac{k|z|+1}{r}\right)^4\right),
\end{equation}
as $r\to 0$
\begin {equation}
\label{}
 h\to e^{i(t-k|z|-1)}c_{02}\frac{k|z|+1}{r},
\end{equation}
as $|z|\to \infty$
\begin {equation}
\label{}
 h\to e^{i(t-k|z|)}\left(c_{02}\frac{k|z|}{r} + c_{01} +
 \frac{3c_{02}}{2}\frac{r}{k|z|}-
 \frac{5c_{02}}{8}\left(\frac{r}{k|z|}\right)^3\right).
\end{equation}
In this case a solution exists containing an arbitrary function
$g$.
\begin {equation}
\label{}
 h=g(t-R)\left(c_{01}+c_{02}\left(\frac{R}{r}+\frac{r}{R}\right)\right).
\end{equation}

\paragraph{Case $\nu=1$}
\begin {equation}
\label{} h=e^{i(t-R)}\left(i+\frac{1}{R}\right)
\left(c_{11}\frac{R}{r}+
c_{12}\left(1-\frac{1}{3}\left(\frac{r}{R}\right)^2\right)\right).
\end{equation}
As $r\to \infty$
\begin {equation}
\label{}
 h\to e^{i(t-r)}\left(i+\frac{1}{r}\right)\left(c_{11}+\frac{2}{3}c_{12}\right),
\end{equation}
as $r\to 0$
\begin {equation}
\label{}
 h\to e^{i(t-k|z|-1)}\left(i+\frac{1}{k|z|+1}\right)\left(c_{11}\frac{k|z|+1}{r}+c_{12}\right),
\end{equation}
as $|z|\to \infty$
\begin {equation}
\label{}
 h\to e^{i(t-k|z|)}\left(ic_{11}\frac{k|z|}{r} + ic_{12} + c_{11}\frac{1}{r}+
 \frac{ic_{11}}{2}\frac{r}{k|z|}+c_{12}\frac{1}{k|z|}\right).
\end{equation}

\paragraph{Case $\nu=2$}
\begin{gather}
\label{}
h=e^{i(t-R)}\frac{R}{r}\left(-1+3i\frac{1}{R}+3\frac{1}{R^2}\right)
\left(c_{21}\left(1-\left(\frac{r}{R}\right)^2\right)^2\right.\notag\\
 +
\left.c_{22}\left(4\frac{r}{R}+6\frac{r}{R}
 \left(1-\left(\frac{r}{R}\right)^2\right)+
 3\left(1-\left(\frac{r}{R}\right)^2\right)^2
 \ln\left(\frac{R+r}{R-r}\right)\right)\right).
\end{gather}
As $r\to \infty$
\begin {equation}
\label{}
 h\to e^{i(t-r)}4c_{22}\left(-1+3i\frac{1}{r}\right),
\end{equation}
as $r\to 0$
\begin {equation}
\label{}
 h\to e^{i(t-k|z|-1)}c_{21}\frac{k|z|+1}{r}\left(-1+3i\frac{1}{k|z|+1}
 +3\frac{1}{(k|z|+1)^2}\right),
\end{equation}
as $|z|\to \infty$
\begin {equation}
\label{}
 h\to e^{i(t-k|z|)}\left(-c_{21}\frac{k|z|}{r}-16c_{22} + 3ic_{21}\frac{1}{r}+
\frac{1}{k|z|}\left(\frac{3c_{21}}{2}r+48ic_{22}+3c_{21}\frac{1}{r}\right)\right).
\end{equation}

Some solutions  asymptotically diverge but linear combinations of
solutions always exist  having any desirable degree of convergence
at infinity. For example  we can linear combine six above
presented solutions into two independent expressions having zero
asymptotics at infinity.

\begin{gather}
 h_5=const\; e^{i(t-R)}\left(\frac{r^2}{R^2} - \frac{r^3}{R^3} +
 {\frac{3i}{R}}\left(1 - 2\frac{r}{R} -
 {\frac{1}{3}}\frac{r^2}{R^2} + \frac{r^3}{R^3}\right)
 \right.\notag\\
 +\left.{\frac{3}{R^2}}\frac{R}{r}\left(1 -
\frac{r^2}{R^2}\right)^2\right).
\end{gather}
 As $r\to \infty$
\begin {equation}
\label{}
 h_5\to -const\; e^{i(t-r)}\frac{i}{r},
\end{equation}
as $r\to 0$
\begin {equation}
\label{}
 h_5\to const\; e^{i(t-k|z|-1)}\frac{3}{(k|z|+1)r},
\end{equation}
as $|z|\to \infty$
\begin {equation}
\label{}
 h_5\to const\; e^{i(t-k|z|)}\frac{3}{k|z|}\left(i +\frac{1}{r}\right).
\end{equation}

And
\begin{gather}
 h_6=const\; e^{i(t-R)}\left(6 - 6
 \frac{r^2}{R^2} -
 \frac{6i}{R}\left(1 + \frac{r^2}{R^2}\right) +
 \frac{6}{R^2}\left(5 - 3
 \frac{r^2}{R^2}\right)\right.\notag \\
 +\left.
\frac{3R}{r}\left(-1+\frac{3i}{R}+\frac{3}{R^2}\right) \left(1-
\frac{r^2}{R^2}\right)^2
 \ln\left(\frac{R+r}{R-r}\right)\right).
\end{gather}
As $r\to \infty$
\begin {equation}
\label{}
 h_6\to -const\; e^{i(t-r)}\frac{12i}{r},
\end{equation}
as $r\to 0$ \begin {equation} \label{}
 h_6\to const\; e^{i(t-k|z|-1)}\frac{12}{k|z|+1}\left(i +\frac{4}{k|z|+1}\right),
\end{equation}
as $|z|\to \infty$
\begin {equation}
\label{}
 h_6\to const\; e^{i(t-k|z|)}\left(\frac{12i}{k|z|} +
  \frac{3}{(kz)^2}\left(16 + r^2\right)\right).
\end{equation}

Obtained solutions describe waves outgoing from an imaginary point
source located outside of the brane. For waves in the regions
$z>0$ and $z<0$ the source is located  at ($r=0, z=-1/k$)  and
($r=0, z=1/k$) respectively.  Notice that in accordance with
\cite{Randall99a}  $1/k$ is of the Planck length order ($\sim
10^{-33}$ sm), so for greater distances, one may suggest that the
source is located on the brane.

Since our solution describes waves propagating with the speed of
the light, we can consider them as gravitational waves in the
5-dimensional spacetime. However since a wave propagates with the
speed of the light in the $R$ direction, the speed of the wave in
the $r$ direction i.e. on the brane must be greater due to the
pure geometrical reason. Indeed, an four-dimensional observer
residing on the brane does not see any tachyons. A wave on the
brane corresponds to variations of a matter distributed on the
brane. In order to imitate a gravitational radiation of a  point
imaginary source in the 5-dimensional spacetime, a variations
phase speed must exceed the speed  of the light.  Notice that
contrary to the group speed the wave's phase speed may be of any
value. At distances $>>1/k$ the difference between the speed of
the light  and the speed of the wave on the brane is $\sim 1/r^2$.

\bigskip
Above we in detail analyzed some solutions which have simple
analytical representations. The general solution \eqref{ansatz}
gives possibilities to study various gravitational fields and
sources. It is demonstrated that perturbations exist which remain
small everywhere in the 5-dimensional spacetime. It signifies that
at this point the RS-metric is not self-contradictory. On the
other hand existing of wave solutions suggests that the brane may
radiate in the outer space and a violation of the brane
energy-momentum conservation may take place. Static solutions may
give some information about possible nonperturbative solutions.

\section*{Acknowledgments}
Author would like to thank I.Ya.Aref'eva, M.G.Ivanov and
I.V.Volovich for fruitful discussions.

\end{document}